\documentstyle[twoside,fleqn,espcrc2]{article}

\def\ben{\begin{equation}}
\def\een{\end{equation}}
\def\bea{\begin{eqnarray}}
\def\eea{\end{eqnarray}}
\input amssym.def
\input amssym.tex

\newcommand{\AmS}{{\protect\the\textfont2
  A\kern-.1667em\lower.5ex\hbox{M}\kern-.125emS}}

\begin{document}
\title{HyperK\"ahler Manifolds \& Multiply-Intersecting Branes}
\author{G. W. Gibbons
\\
D.A.M.T.P.,
\\ Cambridge University, 
\\ Silver Street,
\\ Cambridge CB3 9EW,
 \\ U.K.}
\begin{abstract}
This is a summary of the work in the author's  recent paper
with this title written with Jerome Gauntlett, George Papadopoulos
and Paul Townsend, hep-th/97022012. We showed how to construct
hyper-K\"ahler 8-metrics in terms of arrangements
of three-dimensional hyperplanes in six-dimensional euclidean space.
The slopes of the planes define two relatively prime
integers $(p,q)$.
Under reduction  to ten dimensions and T-duality
we get  a geometric picture of the action of $SL(2,{\Bbb Z})$
on the $(NS \otimes NS ,q R \otimes 
R) $ 5-branes
of type IIB string theory. Configurations are exhibited with 
$3 \over 16$'th SUSY.
\end{abstract}
\maketitle

\section{INTRODUCTION}
The simplifications afforded by going to eleven dimensions
are by now widely appreciated. With this in mind
we construct non-singular solutions
of the equations of motion of 11-dimensional supergravity
taking the form
\ben
ds^2_{11}=H^{ -{2 \over 3}}\eta_{\mu \nu} dx^\mu dx^\nu + Hds^2 _8,\label{metric}
\een
\ben
F_4=\pm { 1\over 6} \epsilon _{\mu\nu \lambda} dx^\mu \wedge
dx ^\nu \wedge dx ^\lambda \wedge d \bigl ({ 1\over H}\bigr ),\label{Ffield}
\een
\ben
\nabla ^2 _8 H=0 \label{brane} .
\een
We could take the Riemannian 8-metric to be merely Ricci flat and we would
then obtain a solution but since we are interested in BPS solutions we shall,
as indicated by our title, choose the metric to be HyperK\"ahler
which will (for an appropriate choice of sign in (\ref{Ffield})
imply that the solution admits Killing spinors.

\subsection{Hyperk\"ahler metrics}
Recall that these are $4k$ dimensional Riemannian metrics $X_{4k}$
which are

\medskip \noindent $(i)$ {\bf Hypercomplex}, i.e. it admit three integrable complex
structures $I,J,K$ satisifying the quaternion algebra $I^2=-1$, $IJ=K$ etc

\medskip \noindent$(ii)$ {\bf Hermitean}, i.e. the complex structures
act as  isometries of the metric: $g(IX,IY)=g(X,Y)$ etc.
equivalently $\Omega_I(X,Y):=g(IX,Y)=-\Omega_(Y,X)$.

\medskip \noindent $(iii)$ {\bf K\"ahler}, i.e. the three two forms $\Omega_I$
are closed, $d {\Omega _I}=0$.

\medskip
It follows that the three 2-forms and the 3 complex structures
are covarariantly constant with respect to the Levi-Civita 
connection $\nabla_g$
of the metric $g$ and that the holonomy group $ Hol (\nabla _g)
\subseteq Sp(k) \subset SO(4k)$. It also follows that
there is {\sl at least}  a $k+1$ dimensional family
of covariantly constant spinors. The number of constant spinors
may exceed $k+1$ if the holonomy group
is a proper subgroup of $Sp(k)$. To count
the constant spinors it suffices to count the number of singlets
in the decomposition of the spinor represention
of $SO(4k)$ with respect to  the subgroup $Hol$.

In fact our particular applications we used 
8-dimensional {\it toric} hyperK\"ahler manifolds.
Then if the harmonic function $H=1$ the holonomy
trivially allows one to read off the 
number of Killing spinors.
Recalling that $Sp(2) \equiv Spin(5)$
and $Sp(1)\equiv Spin(3)\equiv SU(2)$
we have the following possibilities 
for the holonomies and  number of covariantly constant spinors:

\vskip.5cm
\begin{tabular}{|r|c|l|}\hline
Holonomy group & \#  Spinors &  Example\\ 
\hline\hline
${\rm id}$ & 16 &  ${\Bbb E}^8 $ \\
\hline
$Sp(1)$ & 8  & ${\Bbb E}^4 \times $ Taub-NUT \\
\hline
$Sp(1)\times Sp(1) $ & 4 & Taub-NUT $\times$ Taub-NUT \\
\hline
$Sp(2)$ & 3  & Lee-Weinberg-Yi \\  
\hline
\end {tabular}
\vskip .5cm
Our general metric admits no Killing spinors and 
induces the following spinor decomposition
\bea
SO(10,1) &\supset & SO(2,1) \times SO(8):\nonumber \\  {\bf 32}&\rightarrow &({\bf 2}, {\bf 8}_s) \oplus ({\bf 2}, {\bf 8}_c).\nonumber\\ 
\eea 
We  may get 3, 4, or 8 $Spin(2,1)$ doublets respectively according to the following
symmetry breaking patterns:
\bea
SO(8) \subset Sp(2):~~~{\bf 8}s &\rightarrow &{\bf 5}\oplus {\bf 1} \oplus{\bf 1} \oplus {\bf 1} \nonumber \\
{\bf 8}c &\rightarrow &{\bf 4}\oplus {\bf 4} \nonumber \\
\eea
\bea
Sp(2) \subset Sp(1) \times Sp(1):~~~{\bf 5} & \rightarrow & ({\bf 2},{\bf 2}) \oplus ({\bf 1},{\bf 1}) \nonumber \\
{\bf 4} & \rightarrow & ({\bf 2},{\bf 1}) \oplus ({\bf 1},{\bf 2}) \nonumber \\ 
\eea  
\bea
Sp(1) \times Sp(1) \subset Sp(1):~~~{\bf 8}s & \rightarrow & {\bf 5} \oplus {\bf 1} \oplus{\bf 1} \oplus {\bf 1}
\nonumber \\
{\bf 8}c & \rightarrow &  {\bf 4} \oplus {\bf 4} \nonumber \\
({ \bf 2},{\bf 1})& \rightarrow & ({\bf 1},{\bf 1})\oplus ({\bf 1},{\bf 1})\nonumber \\
({\bf 1},{\bf 2}) & \rightarrow & ({\bf 1},{\bf 2}) \nonumber \\
({\bf 2},{\bf 2}) &\rightarrow & ({\bf 1},{\bf 1})\oplus ({\bf 1},{\bf 1}).\nonumber \\
\eea

Thus the fraction of the maximum allowed supersymmetry is given by

\vskip.5cm
\begin{tabular}{|r|c|}\hline
Holonomy group & Fraction of Maximum SUSY\\ 
\hline
${\rm id}$ & 1\\
\hline
$Sp(1)$ & ${ 1\over 2}$ \\
\hline
$Sp(1)\times Sp(1) $ & ${1 \over 4}$ \\
\hline
$Sp(2)$ & ${3 \over 16 }$\\  
\hline
\end {tabular}
\vskip .5cm

The principle  novelty here is the apparently new example of a 
situation with$ 3\over 16$ SUSY.

\section{TORIC HYPERK\"AHLER MANIFOLDS}
This general class of metrics has found a number of physical applications
. By definition they admit  a triholomorphic action
of the $k$-dimensional torus group $T^k \equiv U(1)^k$. 
It turns out that the metric may be written 
in coordinates adapted to the torus action as
\ben
ds^2_{4k}= U_{ij} d{\bf x}^i d{\bf x}^j + (U^{-1})^{ij} (d\phi_i+A_i) 
( d\phi _j +A_j).
\een
and the K\"ahler forms by
\ben
\Omega_I= (d \phi_i + A_i) \wedge dx_1^i -  U_{ij} d x_2 ^i \wedge dx_3 ^j~~~{\rm etc}.
\een
The $k$ Killing fields generating the torus action are
$\partial \over \partial \phi_i$ and clearly for each K\"ahler form
\ben
{\cal L}_{  {\partial \over \partial \phi_i }}   \Omega_I=0.
\een
A useful fact is that the Cartesian coordinates ${\bf x}^i$ are moment maps
(or Hamiltonians) for this torus action.
In addition to the torus action the isometry group contains
an additional $SO(3)$ action.

It is extremely convenient to introduce a privileged
ortho-normal frame $E= ({\bf E}_i, E^i)$ 
by  diagonalizig the matrix $U$:
\ben
U_{ij}=(K^t K)_{ij}
\een
so that
\ben
({\bf E} _i ,  E^j) =( K_{ij}d {\bf x}^j, (d\phi_j+ A_j) K^{ji}).
\een
Thus, in a hopefully obvious notation,:
\ben
\Omega = E^i \wedge {\bf E}_i- {\bf E} \times {\bf E}.
\een
The privileged frame is invariant under the torus action
\ben
{\cal L}_{  {\partial \over \partial \phi_i }} E =0.
\een
Moreover in this frame the  $so(4k)$ Lie
algebra valued   connection
one-forms $\Theta \in sp(k)$.
The spinor covariant $\nabla$  acts on the components $\psi$
of  spinors in the adapted spin frame  as
\ben
\nabla \psi = d \psi  + { 1\over 2} \Theta ^{\mu \nu} \Gamma _{\mu \nu} \psi
\een
Thus, by projecting into the singlet summand under the decomposition
of the spinor 
represenation of $so(4k)$ with respect to its $sp(k)$ subalgebra we obtain
the previously advertized  covariantly constant spinor fields.
These $k+1$  Killing spinors are clearly invariant under the torus action:
\ben
{\cal L}_{  {\partial \over \partial \phi_i }} \psi =0.
\een
The significance of this remark will be apparent later.

\subsection{Lindstr\"om-Roc\v ek-Pedersen-Poon equations}

To specify a toric hyperK\"ahler metric it suffices to give the
matrix $U_{ij}$. The components of the connection
$A_i=\omega _{ik}^r r d x^k_r$ then follow up to gauge
equivalence from the eponymous equations of this subsection
\ben
\partial ^r_j \omega _{ki}~^s -\partial ^s _k \omega _{ji}^r = 
\epsilon ^{rst} \partial ^t_j U_{ki} \label{Poon}.
\een
The integrability conditions: for which are 
\ben
\partial ^t_{[j} U_{k]i}=0\label{curl}
\een
and
\ben
{\partial }_i \cdot {\partial} _j U_{rs}=0\label{harm}.
\een
It is a remarkable fact the the {\sl non-linear}
Einstein equations reduce
to a set of {\sl linear} equations which 
essentially reduce to the requirement the components
of the matrix $U$ are harmonic on euclidean 3-planes.

\subsection{Simple examples}

Let us now turn to the  case
of present interest $k=2$.
We will always choose the two angles $\phi \in(0,2\pi]$.
The simplest examples are well known but
a few points are worth mentioning.
Consider for example the  vacuum or ground state.
This is not altogether trivial
and already we see the modular group $SL(2,{\Bbb Z})$
entering in a natural way as a gauge symmetry.
To specify a flat solution we must give  a constant
 metric, call it $U^{\infty}_{ij}$,
on a 2-torus
or equivalently we must specify
a 2-dimensional lattice. The  basis vectors 
of the lattice make an angle
$\theta$ given by
\ben
\cos \theta = -{  U^ \infty _{12} \over 
\sqrt{ U^\infty _{11} U^\infty _{22} } }
\een
Taking into account the freedom to change basis
we have that the flat metrics  correpond to elements of the double coset 
\ben
SL(2, {\Bbb Z}) \backslash GL(2,{\Bbb R})/SO(2).\een

The next simplest examples are ${\Bbb E}^4 \times$ {\rm Multi-Taub-NUT},
which represent parallel 6-branes in
type of IIA
supergravity in ten dimensions. 
The second factor looks like
\ben
ds^2= H^{-1} (d \phi + {\bf \omega} \cdot d{\bf x} ) ^2 + H d {\bf x} ^2
\een
we have
\ben
\nabla  \times  \omega = \nabla H
\een
and we choose
\ben
H=1+ \sum _{\rm points }   { 1\over 2} { 1 \over |{\bf x}-{\bf a}|} 
\een
As is well known the coordinate singularities
at ${\bf x}={\bf a}$ correpond geometrically to
fixed points of the $T^1\equiv S^1$ action
generated by $\partial \over \partial \phi$.
Absence of singularities forces
the points to be distinct and also fixes the coefficient
of ${\bf x}$ to be the same for all points.
To get down to ten dimensions 
we quotient by this action and regard the points
$\bf a$
as giving the locations of the 6-branes in the
transverse three directions.
It is essential for the physical interpretation
that we preserve supersummetry
and that is why we emphasised the fact that the Killing spinors
were invariant under the torus action. 
Such Killing spinors remain Killing spinors of the reduced theory.

\subsection{Arrangements of Hyperplanes}

The class of toric hyper-K\"ahler metrics
is quite large but it turns out the have a simple geometrical
description in terms of arrangements of hyperplanes. Restricting
ourself to the 8-dimensional case
\ben
U_{ij}= U^\infty_{ij} + \sum _{\rm hyperplanes} { p_ip_j \over 
|p {\bf x}^1 + q {\bf x}^2-{\bf a}| }, 
\een
with $p_1=p$ and $p_2=q$.The fixed point
coordinate singularities are are now located on the three
dimensional hyperplanes in  ${\Bbb E}^6$ given by  
\ben
p{\bf x}^1 + q {\bf x} ^2 =a.
\een
Using the hyper-k\"ahler quotient technique,
which will not be explained here, one discovers
that the hyperplanes will be fixed points
of a smoothly acting  $S^1$ subroup of the
$T^2$ isometry group  if  the quantities specifying the slopes
 $(p,q)$, which also specify the subgroup,
are relatively prime integers. In addition
all planes must be distinct and triple intersections are excluded.

Geometrically it is clear
that the basis chosen allowing us to regard  ${\Bbb E}^6$ 
as ${\Bbb E}^3\oplus {\Bbb E}^3=({\bf x}^1,{\bf x^2})$ is arbitrary
up to  the action of the modular group. 
In our basis $(1,0) \rightarrow $ the ${\bf x}^1$ axis and  
$(2,1) \rightarrow $ the ${\bf x}^2$ axis,
the two be tilted at the angle $\theta$. 
If $\theta= {\pi \over 2}$ the two basis-planes are at right angles
and restrcting $(p,q)$ to these values gives the Multi-Taub-NUT $\times$
Multi-Taub-NUT metrics with holonomy $Sp(1)\times Sp(1)$.
with ${ 1 \over 4}$ SUSY. Keeping the same restriction on $(p,q)$  
but tilting the basis planes  breaks this down to $Sp(1)$ with 
$ { 3 \over 16}$'th  SUSY. If we have just two planes we recover
the Lee-Weinberg-Yi metric which arises as the relative
moduli space of three $BPS$ monopoles in $N=4$ SUSY SU(4)-Yang-Mills
maximally broken to $U(1)\times U(1) \times U(1)$ by a Higgs
in the adjoint representaion. 

\section{THE TYPE IIB VIEWPOINT}

Obviously we can  pick one of the circle subrgroups of  $T^2$ 
and 
reduce
to ten  dimensions to get a solution of the Type 
IIA theory. This solution admits a circle action and so we
T-dualize to get solution  of the Type IIB theory. Otherwise
said, we may descend to nine dimensions where there is no distinction 
between
A and B and and come back up to the IIA theory.
The Type IIB solutions will not necessarilly be non-singular
even though we started in eleven dimensions with non-singular
solutions. The ten-dimensional solution will have a 
Killing vector which we call $\partial \over \partial z$.

The resulting metric is, in Einstein conformal frame,
\bea
ds ^2_{10}&= &
({\rm det}U)^{ 3 \over 4} \Bigl [ ({\rm det}U)^{-1} 
\eta_{\mu \nu} dx ^\mu dx^ \nu \nonumber \\ &+& ({\rm det}U)^{-1} U_{ij} 
d{\bf x}^i \cdot d{\bf x}^j + dz^2 \Bigl ],\nonumber \\
\eea
In addition 
\ben
B_i = A_i\wedge dz
\een
and
\ben
\tau= l+i e^{-\phi}= - 
{ U_{12} \over U_{11} } +i { \sqrt {{\rm det}U} \over U_{11}},
\een
where  $l$ and $\phi$ are respectively
the axion and dilaton
and $B_1$ is the $NS\otimes NS$ and $B_2$ the 
$R \otimes R$  2-form potential. The action of $S\in SL(2, {\Bbb Z})$
associated with the torus under which
\ben
U \rightarrow (S^{-1})^t U S^{-1}
\een
and under which $B_i$ transform as a doublet induces a fractional
linear transformation on $\tau$ via the relation
\ben
{ U \over \sqrt{ {\rm det} U } }= { 1\over {\rm Im}\tau} 
\pmatrix{
1&-{\rm Re} \tau\cr
-{\rm Re}\tau & |\tau|^2 \cr
}. 
\een
Of course, as far as the classical Type IIB theory is concerned,
we have an action
of $SL(2,{\Bbb R})$ on the classical solutions
but if we lift this to eleven dimensions
it will in general take non-singular solutions to singular solutions.
In the Type IIB theory, in which the classical solutions are in general
singular,  one usually restricts this $SL(2,{\Bbb R})$ action
to
$SL(2,{\Bbb Z})$ by appealing to quantum effects. The striking
thing about our work is that in eleven
dimensions is that this restriction arises from a demanding
the classical solutions be regular. 

\subsection{Examples}
The discussion of the previous section may be fleshed out by
looking at the simplest example. We set $\theta= {\pi \over 2}$
and thus
\ben
U= \pmatrix{
H_i({\bf x}^1)& 0\cr
0& H_2({\bf x}^2)\cr
}
\een
where $H_1$ and $H_2$ are harmonic functions on ${\Bbb E}^3$.
Since
\bea
ds^2 _{10}& =& (H_1 H_2)^{ 3\over 4} \Bigl [ { 1\over H_1 H_2} ( -dt^2 + \eta _{\mu \nu} d x^ \mu d x^ \nu \nonumber \\& + &{ 1\over H_2} d {\bf x}^1 \cdot d{\bf x}^1   + { 1\over H_1} d{\bf x}^2 \cdot d{\bf x}^2  + d z^2 \Bigr ].\nonumber \\
\eea 
This is readily recognized as the metric commonly referred to
as 
the orthogonal intersection of a collection
of parallel $NS\otimes NS$ and $R\otimes R$
5-branes on a set of two-branes. The $NS\otimes NS$
 brane coordinates are $(x^\mu, {\bf x}^1)$, 
$R\otimes R$ coordinates are $(x^\mu, {\bf x}^2)$
and the 2-brane coordinates are $x^\mu$
where now, since we are in 10 dimensions, the latin
indices run from 0 to 1.
Since the metric is independent of the 
mutally transverse coordinate $z$ the 5-branes are delocalized
or \lq stacked \rq in that direction.

If we  may now pass to the general case when the planes are tilted
we see that it is reasonable to interpret the general
hyperplane $p{\bf x}^1 + q {\bf x}^2 ={\bf a}$ as a Type IIB
5-brane with charge $(p,q)$ with $ 3\over 16$ 'th SUSY.

\subsection{Non-Orthogonal D-branes}

By dualizing the Type IIA solutions in a different direction
one may obtain a solution contining only D-five branes.
If $X ^i=({\bf x}^i, \phi _i)$ one has, again in Einstein conformal gauge
\bea
ds^2_{10} &= & ({\rm det} U)^ { 1\over 4} \Bigl [- dt^2 +  (d x^1)^2  \nonumber \\&+&U_{ij} dX^i dX^j ]\nonumber \\
B_2 & =& A_i d \phi _i \nonumber \\
\tau &= &i\sqrt {{\rm det}U}\nonumber .\\ 
\eea
The two branes intersect on a string
extended along the $x^1$ direction.
thus 
if $\theta=0$ one 5- brane occupies the $12345$ directions
and the other the $16789$ directions. If $\theta \ne 0$
the $2345$ directions and $6789$ directions are
rotated with respect to each other at an angle by
an $SO(8)$  element $O$ which is block diagonal
in the $2-6$, $3-7$, $4-8$ and $5-9$ 2-planes,
each block looking like\ben
\pmatrix {
\cos& \sin \theta \cr
-\sin \theta & \cos \theta  \cr 
}.
\een
Using quaternion notation with 
\bea 
& & {\Bbb E}^*=(X^1,X^2) 
\equiv {\Bbb H}^2 =\nonumber \\& & (x^2 + i x^ 3 + j x^4 + k x^5, x^6 +i x^7 + j x^8  + k x^9 ),\nonumber
\\
\eea
 we see that one may think of $O$ as lying
in $Sp(2) \subset SO(8)$.

Now a D-brane occupying the $1,2,3,4,5$ directions is invariant under
supersymmetries generated by the Type IIB chiral spinors $\epsilon ^i$
satisfying
\ben
\Gamma _{012345} \epsilon^1 = \epsilon ^2.
\een
If the other D-branes were orthogonal it would be invariant under supersymmetries
satisfying
\ben
\Gamma _{016789} \epsilon^1 = \epsilon ^2.
\een
The set of common solutions would be 4 dimensional.
However if they are at an angle one has instead
\ben
R^{-1} \Gamma _{016789}R \epsilon^1 = \epsilon ^2,
\een
where $R(\theta)$ is the  lift to $Spin(8)$
of the $SO(8)$ rotation $O$. Explicitly
\ben
R(\theta) =\exp \{ -{ 1\over 2} \theta ( \Gamma _{26} +\Gamma_ {37}+ \Gamma_ + \Gamma _{48} +\Gamma_{59}) \}
.
\een
After some Clifford algebra one finds that
if $\theta \ne 0$ there are just 3  mutual solutions as expected.

\subsection{Hanany-Witten type solutions}

Using different reductions and different T-duality maps one may 
obtain a great variety of other
solutions with $ 3\over 16$ 'th SUSY. These include solutions
with just $NS\otimes NS$ or just $R\otimes R$ 5-branes
intersecting at an angle. As we have just  have checked 
in the $R\otimes R$ case,
the amount of SUSY is consistent with the stringy analysis
of Dirichlet-5-branes. Perhaps more interesting
are the solutions which, if they were not delocalized in the $z$
direction, would correspond to the configurations
used by Hanany and Witten in their analysis of 
2+1 dimensional  gauge theories on the intersetion of 
$NS \otimes NS$ and $R \otimes R$ 5-branes. In order to
localize the  2-branes we need to solve the equation \ref{brane}
for the harmonic function which gives the 4-Form \ref{Ffield}
appearing in the general metric (\ref{metric}).
In the toric case  when we also assume that $H$
is $T^2$ invariant, this becomes
\ben
U^{ij}\partial_i \cdot \partial _j H=0.
\een
Rather remarkably this equation is  additively separable:
it admits solutions of the form
\ben
H= H_1({\bf x}^1) + H_2({\bf x}^2)\label{2-brane}.
\een

\section{ HOLOMORPHIC CYCLES}

As an illustration
of the  utility of the eleven dimensional
viwe point it is worth pointing
out that the holomorphic geometry of toric hyperK\"ahler manifolds
makes it almots trivial to
construct an interesting class of \lq test probes \rq 
in these geometries. Consider one of the 2-sphere`s worth
of complex structures specified by a unit 3-vector ${\bf n}$
\ben
I_{\bf n}=n_i I + n_2 J + n_3 K.
\een
It determines a direction in each  of the $k$ ${\Bbb E}^3$
factors in the quotient $X_{4k}/T^k$ manfold. Now 
consider a $k$-plane $\Pi \subset {\Bbb E}^{3k} $
containing these directions. Using the
torus action it may be lifted up to $X_{4k}$
to give a $2k$-dimensional submanifold
which one may easily y check is holomorphic
with respect to the complex structure $J{\bf n}$.
By picking the $k$-plane $\Pi$ appropriately
one may obtain in this way a variety of different types
of holomorphic cycles with different topologies.
By Wirtinger`s theorem they are all minimal.
One may also check that they remain minimal
when the extra harmonic functions (\ref{2-brane}) are
included.

\section{CONCLUSION \& PROSPECTS}
Perhaps the most  striking thing about 
our analysis is how simpy 
one may construct extremely elaborate 
non-singular intersecting
brane solutions with modest amounts of supersymmetry
by considering arrangemenst of rational  hyperplanes in six dimensions
and how these give a purely classical geometrical
insight
into the what in Type IIB theory
is thought of as the quantum mechanical breaking of the
classical $SL(2,{\Bbb R})$ down to $SL(2,{\Bbb Z})$.
It was gratifying to see that our analysis of the
amount of supersymmetry is consistent with
that given string theory using by D-brane techniques. 

Given our construction of Hanany-Witten type solutions
it would be interesting to investigate what $N=3$,
2+1 dimensional gauge theories may arise on the intersections.

In the talk I also mentioned the fact that
that it would be interesting to explore further
an aspect of this work which plays an essential
role in the calculations: the fact that
T-duality has, in some sense, the effect of interchanging 
hyperK\"ahler (HK) geometry
with what is sometimes called HKT geometry. This has now been
completed \cite{GPS}. 

In the bibliography below I have restricted myself 
to the paper \cite{GGPT} of which this talk is a summary and the newer
paper \cite{GPS} mentioned above. A complete list of references 
to the original literature related to 
the material discussed above may be found there. 

\section{ACKNOWLEDGEMENT}

It is my pleasant duty to thank the organizers
of SUSY 97 for
the opportunity to visit Philadelphia and to participate
in a very stimulating and enjoyable meeting.


\begin{thebibliography}{99}

\bibitem{GGPT} G Gauntlett, G. W. Gibbons, G Papadoupolus
and P K Townsend, HyperK\"ahler manifolds and multiply intersecting
branes, {\it Nucl.Phys.} {\bf B} in press, hep-th/97012112  

\bibitem{GPS} G. W. Gibbons, G Papadopoulos and K.S. Stelle,
HKT and OKT Geometries on Black Hole Soliton Moduli Spaces,
hep-th/9706207  

\end{thebibliography}
\end{document}